\documentclass[%
 aip,
 amsmath,amssymb,
 reprint,%
]{revtex4-1}

\usepackage{graphicx}
\usepackage{dcolumn}
\usepackage{bm}
\usepackage[utf8]{inputenc}
\usepackage[T1]{fontenc}
\usepackage{mathptmx}
\usepackage{etoolbox}
\usepackage[normalem]{ulem}
\usepackage{indentfirst}
\usepackage{epsfig}
\usepackage{mathrsfs}
\usepackage{amsmath}
\usepackage{nccmath}
\usepackage{float}
\usepackage{color}
\usepackage{ulem}
\usepackage{amssymb}
\usepackage{epstopdf}
\usepackage{dcolumn}
\usepackage{bm}
 \usepackage{graphicx}
 \usepackage{braket}
\usepackage{braket}
\newcommand{\RN}[1]{\textup{\uppercase\expandafter{\romannumeral#1}}}
\newcommand{\bc}{\begin{center}}
\newcommand{\ec}{\end{center}}
\newcommand{\beq}{\begin{equation}}
\newcommand{\eeq}{\end{equation}}
\newcommand{\bea}{\begin{array}}
\newcommand{\eea}{\end{array}}
\newcommand{\bey}{\begin{eqnarray}}
\newcommand{\eey}{\end{eqnarray}}

\makeatletter
\def\@email#1#2{%
 \endgroup
 \patchcmd{\titleblock@produce}
  {\frontmatter@RRAPformat}
  {\frontmatter@RRAPformat{\produce@RRAP{*#1\href{mailto:#2}{#2}}}\frontmatter@RRAPformat}
  {}{}
}%
\usepackage[colorlinks=true,linkcolor=blue, allcolors=blue]{hyperref}%

\makeatother
\begin{document}

\preprint{AIP/123-QED}

\title{Coherent control of the Goos-H\"{a}nchen shift in Otto structure}

\author{Magdi Bajusair}
\affiliation{Department of Physics and Astronomy, College of Science, King Saud University,
P. O. Box 2455, Riyadh 11451, Saudi Arabia}

\author{Mohammad H. Alhakami}
\affiliation{Department of Physics and Astronomy, College of Science, King Saud University,
P. O. Box 2455, Riyadh 11451, Saudi Arabia}

\author{Saeed Asiri}
\altaffiliation{sasiri@kacst.gov.sa} 
\affiliation{Institute of Quantum Technologies and Advanced Computing, KACST, Riyadh 11442, Saudi Arabia}

\date{\today}

\begin{abstract}
We investigate controlling the lateral Goos-Hänchen (GH) shift for a TM-polarized field reflected from Otto structure containing four level N-type atomic medium. The N-type atomic configuration can be formed by coupling the standard three-level $\Lambda$ system to an additional upper energy level through a coherent driving field. The medium can then be switched from transparent to absorptive under the effect of the driving field. In the Otto structure, an air gap typically separates a dielectric prism from a metal film. We show that the sign and magnitude of the GH shift can be highly controlled when the air gap is replaced by the coherent atomic medium. This can be achieved by adjusting the strength of the applied fields to the atomic medium, while the geometrical characteristics of the proposed structure are unchanged.
\end{abstract}

\maketitle

\section{Introduction} 
Goos-Hänchen (GH) shift is a phenomenon that refers to the lateral displacement that appears when an incident light with a finite width undergoes total internal reflection at the interface between optically denser and thinner media. In 1947, Goos and Hänchen carried out the first experimental demonstration of this phenomenon at a glass-air interface \cite{01}. Following this experiment, Artmann explained the GH effect theoretically \cite{02} and provided a well-known formula to calculate the GH shift using the acquired phase shift upon total internal reflection.

GH shift has potential applications in different directions such as sensing \cite{03, 04, 05}, optical switching \cite{06, 006}, and Material characterization \cite{07}. Motivated by these promising applications as well as the experimental achievements in observing the GH shift in different systems \cite{Exp01, Exp001, Exp1, Exp2, Exp3}, investigations of the GH shift have been extended to more sophisticated structural interfaces and schemes. In the recent years, GH shift has been extensively explored in structures such as dielectric slabs \cite{Lai:02, 08, 09}, different layered configurations \cite{L1, L2, L3, L5, L6, L7}, and photonic crystals \cite{10}. Different kinds of materials have been considered in GH shift research for different purposes such as graphene \cite{gr1, 11}, Epsilon-near-zero (ENZ) materials \cite{ENZ1, ENZ2}, Left-handed materials \cite{Left1, Left2}, Weyl semimetals \cite{weyl1, weyl2}, and topological insulators \cite{topological1, topological2, topological3}.

In the previous studies, it is generally not possible to tune and manipulate the GH shift while the actual proposed structure is fixed. To address this limitation, researchers proposed the idea of using quantum coherence effects \cite{CE1, Scully_Zubairy_1997, CE2, CE3} in different atomic systems in which the optical response of these systems can be adjusted and controlled by external fields. For example, Wang et al. proposed a fixed cavity configuration to control the GH shift using a two-level atomic medium placed inside the cavity \cite{c1}. By applying an external control field on the atomic medium to tune its optical response, positive and negative GH shifts were achieved. Following this proposal, researchers explored quantum coherence effects in other atomic configurations to exploit their tunable optical responses to effectively control and manipulate the GH shift \cite{c1, c2, c3, c4, c5, deng2012enhancement, c6, c7, c8, c9, c10}.

However, since the GH shift is usually in a wavelength or even sub-wavelength scale, significant number of research proposals have been primarily focused on enhancement of the phenomenon \cite{Exp2, Lai:02, 09, L3, 10, Left2, c10, enhancement1, GH_SPR1}, making it possible to be measured experimentally for useful applications. Surface plasmon resonance (SPR) structures have been proposed as practical structures to produce large positive and negative GH shifts \cite{enhancement1, GH_SPR1, GH_SPR2}. For example, in Kretschmann–Reather configuration, large positive and negative GH shifts were experimentally detected by exciting SPR at a metal-air interface \cite{GH_SPR1}. Another three-layer SPR system involving four-level atomic system has been recently shown to produce giant positive and negative GH shifts \cite{GH_SPR3}. More recently, a coupler-free SPR system containing a three-level atomic medium were shown to amplify and control the GH shift \cite{GH_SPR4}.

On the other hand, the Otto configuration \cite{Otto}, which is typically made of a dielectric prism and a metal film separated by an air-gap, is used for the excitation of surface plasmon waves at the air-metal interface. In 1978, Lukosz et al. \cite{Lukosz:78} showed that in Otto configuration, there exist two angles of incidence, i.e., $\theta_1$ and $\theta_2$, for which the reflectivity for a TM-polarized wave can be zero. The angles of incidence $\theta_1$ and $\theta_2$ are associated with specific thicknesses of the air-gap, i.e., $d_1$ and $d_2$, respectively. The first zero reflectivity which occurs at $\theta_1$ and $d_1$, is found to be very close to the SPR angle. The second zero reflectivity is located at $\theta_2$ with $d_2$ appears near grazing incidence. The GH shift around these reflectivity dips can become large due to the nature of phase variation near $\theta_1$ and $\theta_2$. The behavior of the GH shift for a TM-polarized field reflected from the conventional Otto structure has been investigated near the first and second reflectivity zeros in \cite{Otto_air1, Otto_air2} and \cite{Otto_air3}, respectively. Thus, the choice of the air-gap thickness in both cases greatly affects the features of the GH shift. In this paper, we investigate the GH shift of the TM-polarized light beam reflected from a three-layer Otto structure containing N-type atomic medium \cite{N_atom1, N_atom2, N_atom3, N_atom4}. We focus on the first zero reflectivity, and show that the GH shift can be coherently controlled near the angle of incidence $\theta_1$ and the optimal air-gap thickness $d_1$. Owing to the modified optical properties of the coherent medium, the air-gap optimal parameters $\theta_1$ and $d_1$ can be shifted from their original values. We show that the GH shift is greatly enhanced around the new optimal conditions associated with the permittivity of the N-type atomic medium $\varepsilon_{2}$. 

\section{Model}
We consider a TM-polarized light beam is incident upon a three-layer Otto structure as illustrated in Fig. 1(a). The proposed Otto structure to control the GH shift of the reflected light contains a glass prism with permittivity $\varepsilon_{1}$, a central layer of thickness $d$ filled with the N-type atomic medium with permittivity $\varepsilon_{2} = 1 + \chi$, and a metal film such as silver with permittivity $\varepsilon_{3}$. As the permittivity of the atomic medium depicted in Fig. 1(b) can be coherently controlled by the applied coupling and driving fields \cite{N_atom1, N_atom2, N_atom3, N_atom4}, it is predicted that the GH shift in Otto structure can be manipulated and controlled by modifying $\varepsilon_{2}$. The N-type atomic configuration can exist for example in Sodium with the D2 line transition at a wavelength of $589.1 \, \text{nm}$. The complex permittivity for silver in this case is about $\varepsilon_{3} = -13.3+0.883i$.

When the incident field has sufficiently large finite width, the GH shift of the reflected light beam can then be calculated using the stationary phase theory \cite{02}, which is expressed as
\begin{equation}
\label{stationary}
S_{r} = - \dfrac{1}{k \sqrt{\varepsilon_{1}} \cos \theta} \frac{d \phi_{r}}{d \theta},
\end{equation}

 \begin{figure}[t]
 \centering
  \includegraphics[width=44mm]{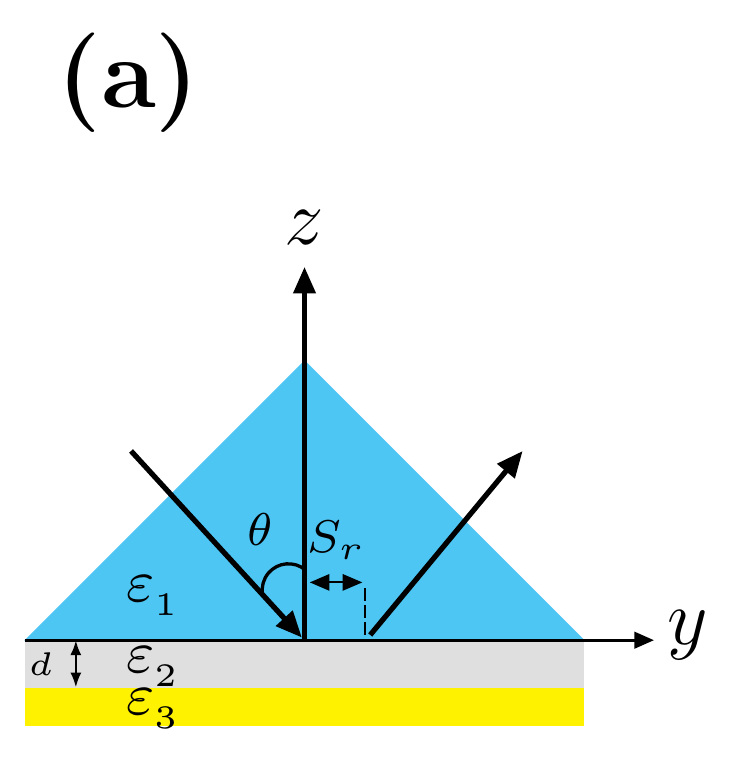} \includegraphics[width=40mm]{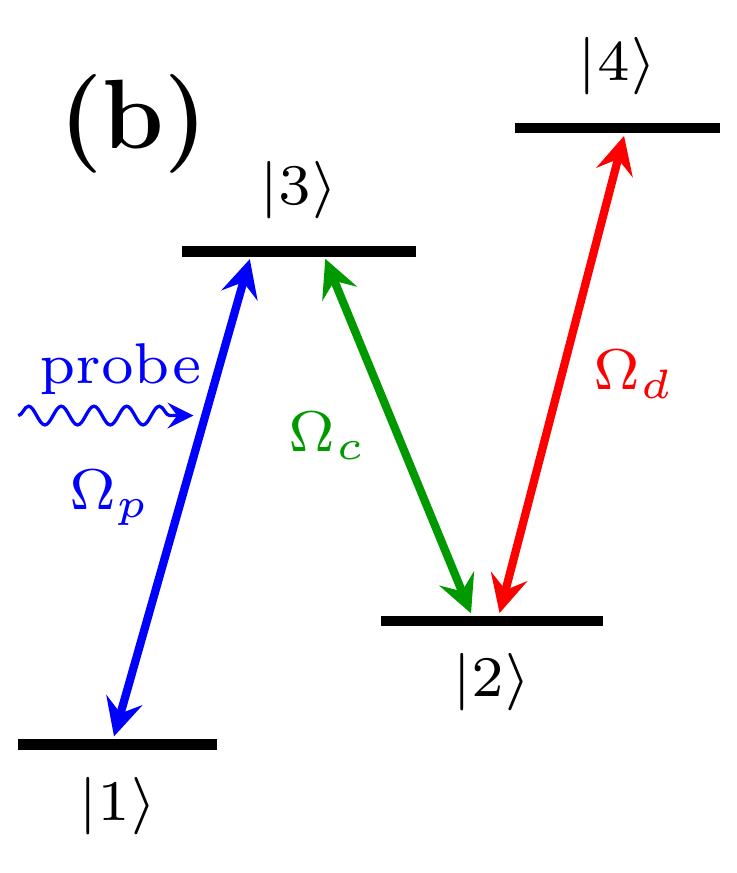}
 \caption{ (a) Schematic of Otto configuration containing three layers, where an atomic medium of thickness $d$ separates a glass prism and a metal film. (b) Energy levels structure of the N-type atomic system.}
 \end{figure} 

where $k = 2 \pi/\lambda_{p} = \omega_{p}/c$ is the wave number for the incident probe field with a wavelength $\lambda_{p}$ and angular frequency $\omega_{p}$, whereas $c$ denotes the speed of light in vacuum. $\phi_{r}$ is the phase shift of the reflected light with reflection coefficient $r(\theta, \omega_{p}, d) = |r(\theta, \omega_{p}, d)| e^{i \phi_{r}}$. For the three-layer structure depicted in Fig. 1(a), the reflection coefficient is defined as 
\begin{equation}
\label{r}
r(\theta, \omega_{p}, d) = \dfrac{r_{12} + r_{23} \; e^{2i k_{z2} d}}{1 + r_{12}  r_{23} e^{2i k_{z2} d}},
\end{equation}
where $k_{z2}$ denotes the $z$ component of the wave vector in the central layer, $r_{ij}$ is the Fresnel reflection coefficient for the TM-polarized light at the interface of medium $i$ and medium $j$, and is defined as 
\begin{equation}
\label{rTE}
r_{ij} = \dfrac{\varepsilon_{j} k_{zi} - \varepsilon_{i} k_{zj}}{ \varepsilon_{j} k_{zi} + \varepsilon_{i} k_{zj} }.
\end{equation}
The $z$ component of the wave vector in the above equations is given as $k_{zi} = k \sqrt{\varepsilon_{i} - \varepsilon_{1} \sin^2 \theta}$. 
\section{N-type atomic medium}
The N-type atomic system \cite{N_atom1, N_atom2, N_atom3, N_atom4} consists of two ground states $\ket{1}$ and $\ket{2}$, and two excited states $\ket{3}$ and $\ket{4}$. The energy level diagram of the N-type atomic system considered in this work is depicted in Fig. 1(b). A weak probe field of angular frequency $\omega_{p}$ and Rabi frequency $\Omega_{p}$ is applied to the transition $\ket{1} \leftrightarrow  \ket{3}$. The transition $\ket{2} \leftrightarrow  \ket{3}$ is coupled by a strong coupling field of angular frequency $\omega_{c}$ and Rabi frequency $\Omega_{c}$. The transition $\ket{2} \leftrightarrow  \ket{4}$ is coupled by a driving laser field of angular frequency $\omega_{d}$ and Rabi frequency $\Omega_{d}$. We assume that the spontaneous decay rate from level $\ket{i}$ to level $\ket{j}$ is denoted by $\Gamma_{ij}$ with $(i, j = 1-4)$. The decoherence between the ground states $\ket{1}$ and $\ket{2}$ is set to be $\gamma_{21}$. 
The interaction picture Hamiltonian describing the N-type atomic system in the dipole and rotating wave approximations can be expressed as
\begin{align}
\label{Hamiltonian}
H &=  -\hbar \big[ (\Delta_p - \Delta_c) \ket{2} \bra{2} + \Delta_p  \ket{3} \bra{3} + (\Delta_p - \Delta_c+\Delta_d) \ket{4} \bra{4} \big] \nonumber \\
 & - \hbar \left( \Omega_p \ket{3} \bra{1} + \Omega_c \ket{3} \bra{2} + \Omega_p \ket{4} \bra{2}  + H.c. \right) ,
 \end{align}
where $\Delta_p = \omega_p - \omega_{31}$ is the probe field detuning and $\Delta_c = \omega_c - \omega_{32}$ is the coupling field detuning, whereas $\Delta_d = \omega_d - \omega_{42}$ represents the driving field detuning, with $\omega_{31}$, $\omega_{32}$, and $\omega_{42}$ denoting the frequencies of the transitions $\ket{1} \leftrightarrow  \ket{3}$, $\ket{2} \leftrightarrow  \ket{3}$, and $\ket{2} \leftrightarrow  \ket{4}$, respectively. 

Equations of motion for the density matrix elements can be derived directly using Liouville equation of motion for the density matrix \cite{Scully_Zubairy_1997} along with the Hamiltonian Eq. \eqref{Hamiltonian}.

Assuming that the probe field is weak, linear response of the atomic medium can be obtained by solving these set of equations \cite{PhysRevA.101.033837} for the coherence element ${\rho}_{31}$ up to the first order approximation in steady state. The linear susceptibility of the atomic medium \cite{GH_SPR3} can then be calculated as
\begin{equation}
\label{chi}
\chi = -\beta \dfrac{d_{21} d_{41} - \Omega_{d}^{2}} {d_{31} (d_{21} d_{41} - \Omega_{d}^{2}) - d_{41} \Omega_{c}^{2}},
\end{equation} 
where $d_{21} = (\Delta_{p} - \Delta_{c}) + i \gamma_{21}$, $d_{31} = \Delta_{p} + i \Gamma_{3}/2$, $d_{41} = (\Delta_{p} - \Delta_{c} + \Delta_{d} ) +  i \Gamma_{4}/2$, and $\beta=N \wp_{13}^{2}/2\varepsilon_{0}\hbar$ with $N$ is the number density of the atomic gas medium and $\wp_{13}$ is the matrix element of the electric dipole moment of the $\ket{1} \leftrightarrow  \ket{3}$ transition. $\Gamma_3$ and $\Gamma_4$ are defined such that, $\Gamma_{3}=\Gamma_{31} + \Gamma_{32}$ and $\Gamma_{4}=\Gamma_{41} +\Gamma_{42} $. The linear susceptibility Eq. \eqref{chi} indicates that the permittivity of the atomic medium $\varepsilon_{2} = 1 + \chi$ can be effectively tuned by modifying Rabi frequencies of the coupling and driving fields as well as the detunings of the applied probe, coupling, and driving fields. 
When the driving field is switched off, i.e., $\Omega_{d} = 0$, and the strong coupling field is resonantly interacting with the transition $\ket{2} \leftrightarrow  \ket{3}$, the probe field absorption vanishes at $\Delta_{p} = 0$, and the absorption profile presents two peaks located around $\pm\Omega_c$. This is the typical linear response of a three-level $\Lambda$ atomic system where a transparency window is observed at $\Delta_{p} =0$. However, when the driving field $\Omega_{d}$ is applied to the $\ket{2} \leftrightarrow  \ket{4}$ transition, an absorption peak is observed at $\Delta_{p} =0$, which can be explained in the dressed-state picture as a two-photon absorption process \cite{N_atom2}. Therefore, the optical response of the N-type atomic medium can be highly tuned by modifying the strength of the coupling and driving fields $\Omega_{c}$ and $\Omega_{d}$, respectively.
 
\section{air gap case} 
We start with the simplest case where the central layer contains an air, i.e., $\varepsilon_{2} = 1$. We present the results of the GH shift for this case in order to see the influence of the coherent medium on the GH shift in the next section in comparison to the air case. When the angle of incidence is greater than the critical angle, total internal reflection occurs at the prism-air interface and the evanescent wave in this case passes through this interface into the air gap. The evanescent wave can then excite the surface plasmon waves on the metal surface which can lead to surface plasmon resonance (SPR) manifested by a sudden dip in the reflectivity profile. Now, as indicated in the introduction, we concentrate on the first zero reflectivity in the three-layer Otto structure.  
Eq. \eqref{r} implies that exact zero reflectivity requires the numerator to be equal to zero, i.e., $r_{12} + r_{23} \; e^{2i k_{z2} d} = 0$. This condition can be solved to find the thickness and angle of incidence at which the first zero reflectivity is achieved. The first zero reflectivity is found to be at $\theta_{1} \approx 43.8^{\circ}$ with $d_1 \approx 653 \, \text{nm}$ $\simeq 1.108 \lambda$, which is very close the SPR angle $\theta_{\text{SPR}} = 43.88^{\circ}$. The reflection phase around $\theta_1$ undergoes sharp change which consequently leads to enhancement of the GH shift around the optimal values $\theta_1$ and $d_1$. Fig. 2 shows the reflectivity and the GH shift for the reflected TM-polarized field as a function of the angle of incidence for several values of the air-gap thickness. It can be obviously seen that the air-gap thickness can determine the direction and magnitude of the GH shift. 
\begin{figure}[htbp]
 \centering
  \includegraphics[width=42mm]{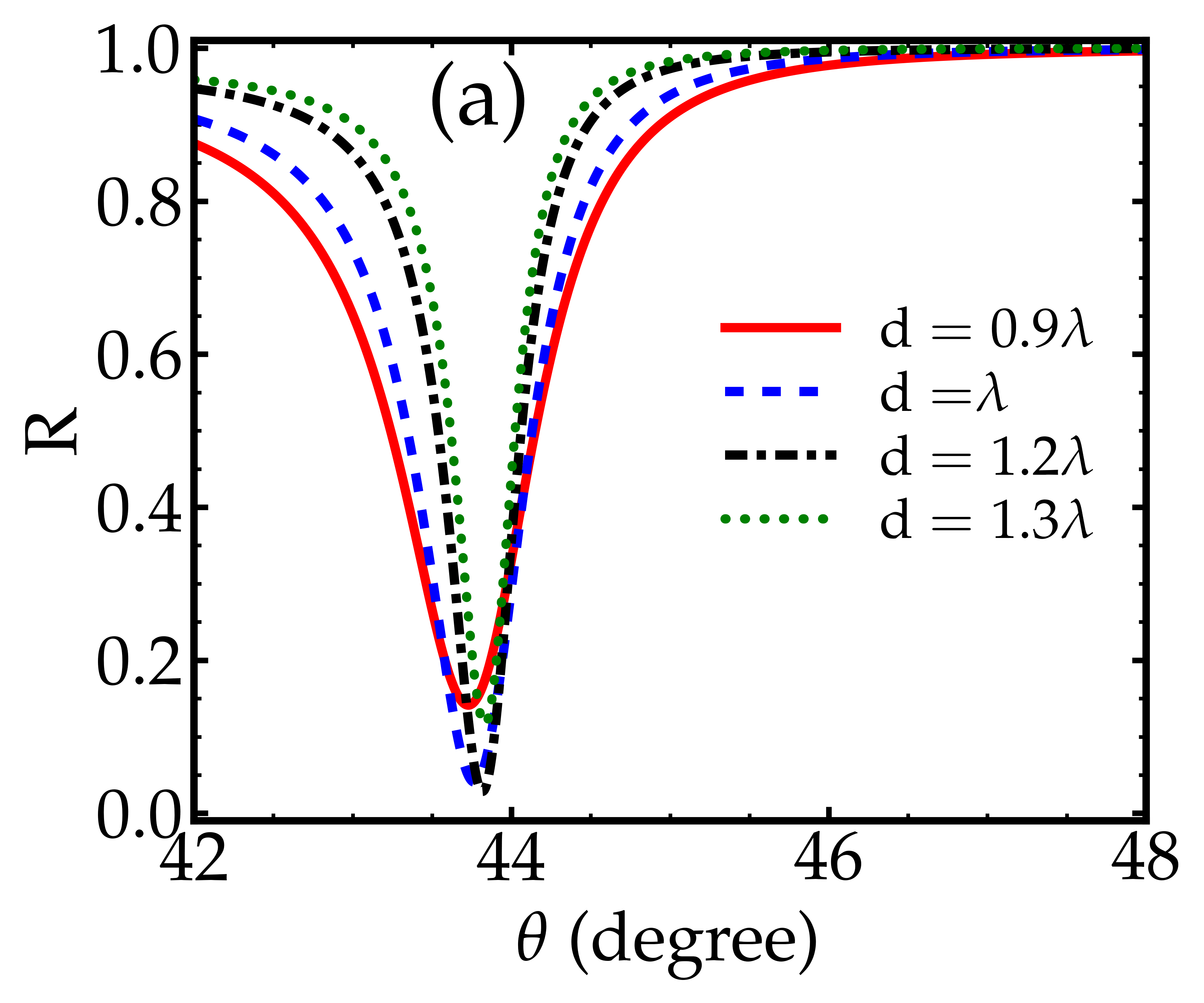} 
  \includegraphics[width=42mm]{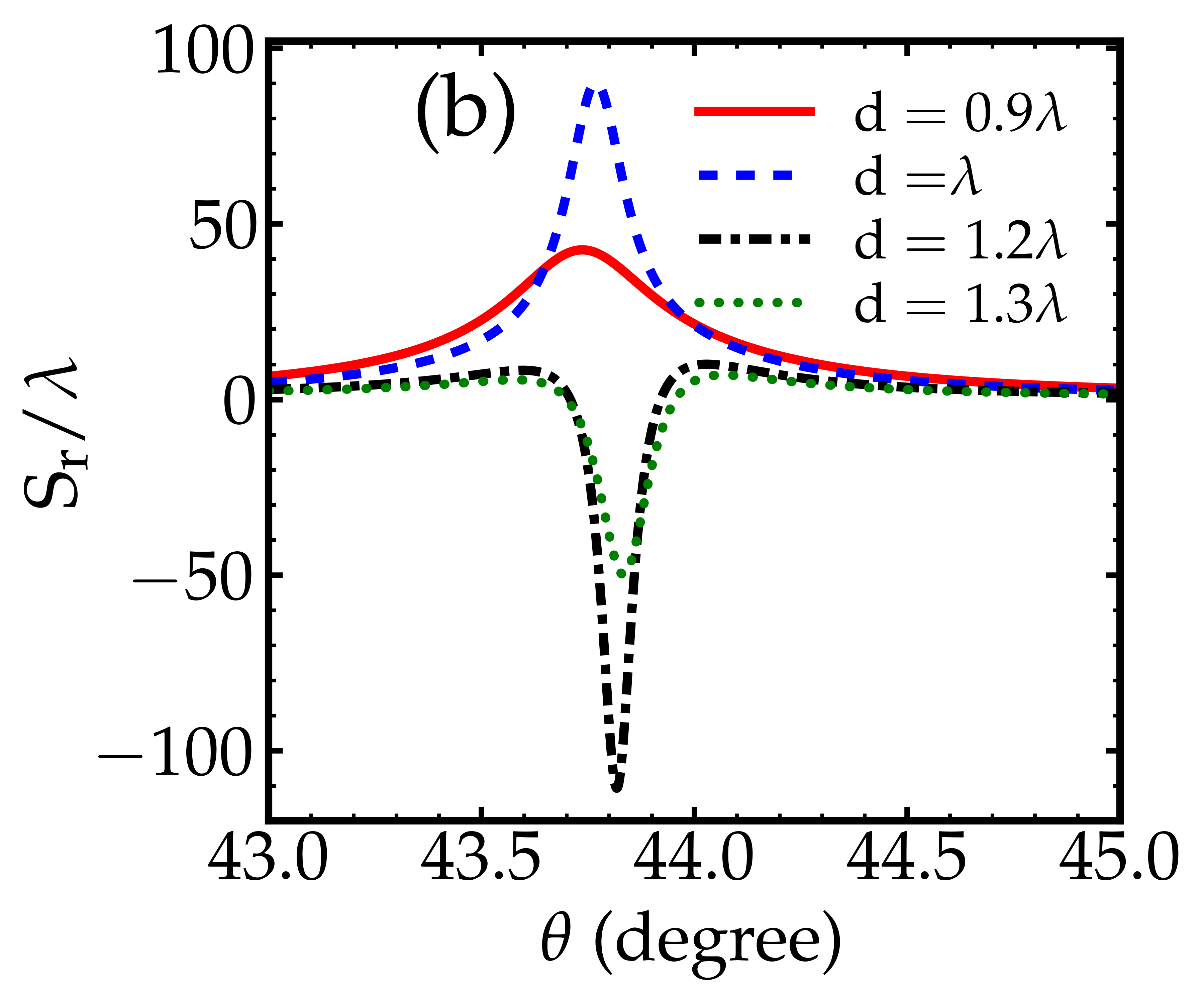}
 \caption{ (a) Reflectivity and (b) GH shift versus the angle of incidence of the reflected TM-polarized light beam from the three-layer structure for different values of the air-gap thickness. Other parameters are $\varepsilon_{1} = 2.25$, $\varepsilon_{2} = 1$, and $\varepsilon_{3} = -13.3+0.883i$}
 \end{figure} 
When $d$ is close to the optimal thickness $d_1$, the reflectivity minimum approach zero, and the GH shift is clearly enhanced. It is also noted that the direction of the GH shift switches from positive to negative when $d$ is below and above the optimal thickness $d_1$, respectively.

\begin{figure}[H]
 \centering
 \includegraphics[width=42mm]{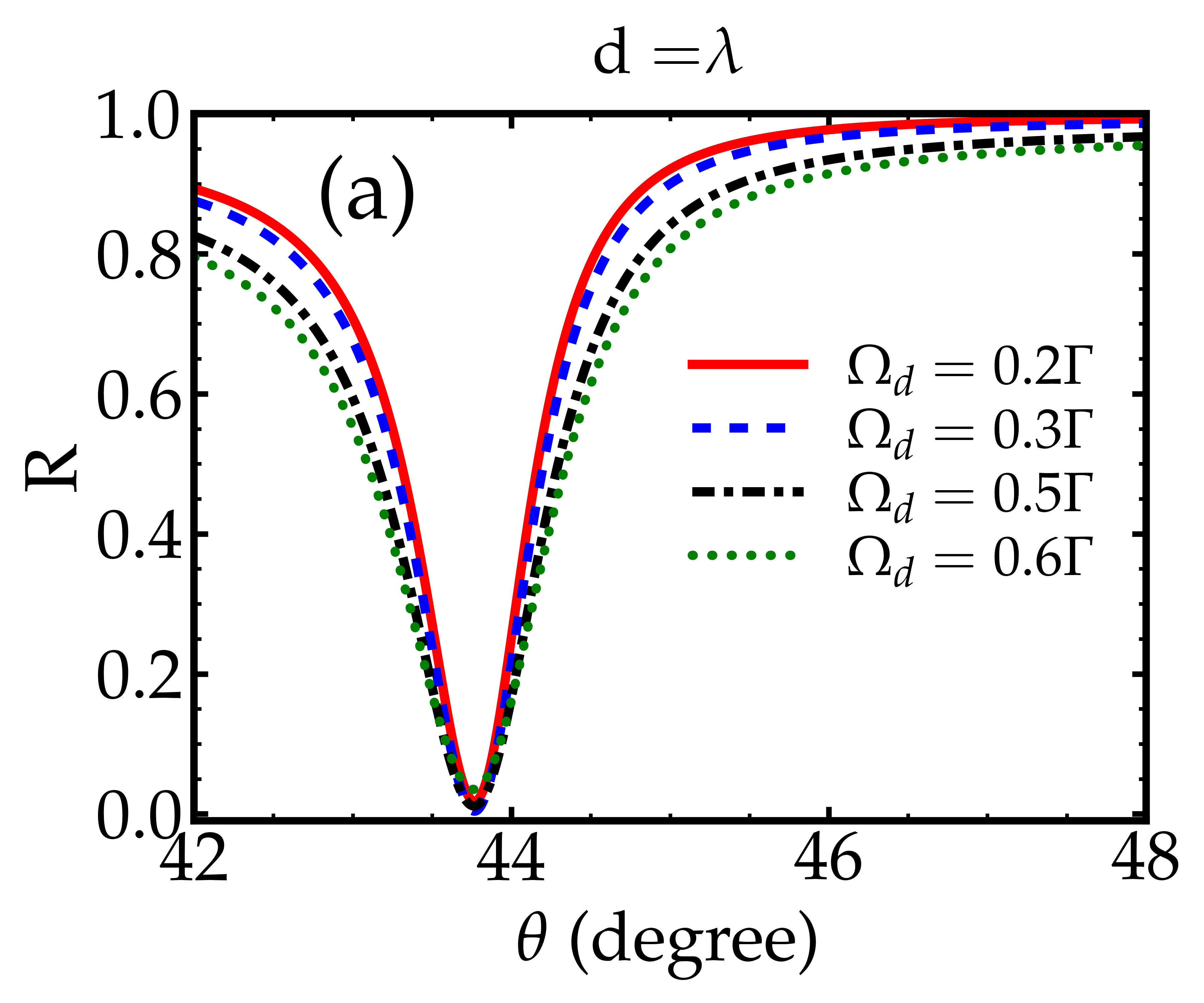} \includegraphics[width = 42mm]{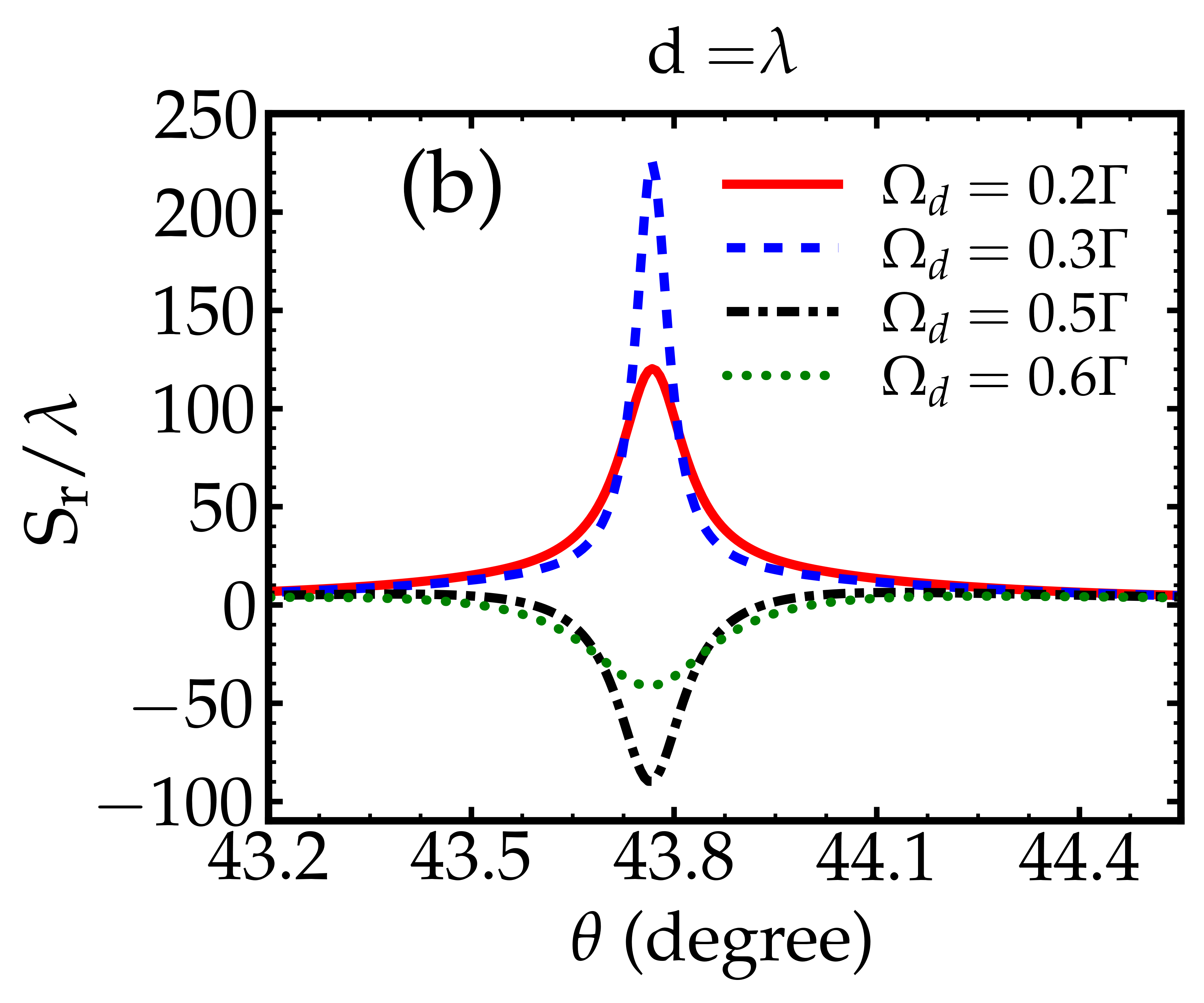} \includegraphics[width=42mm]{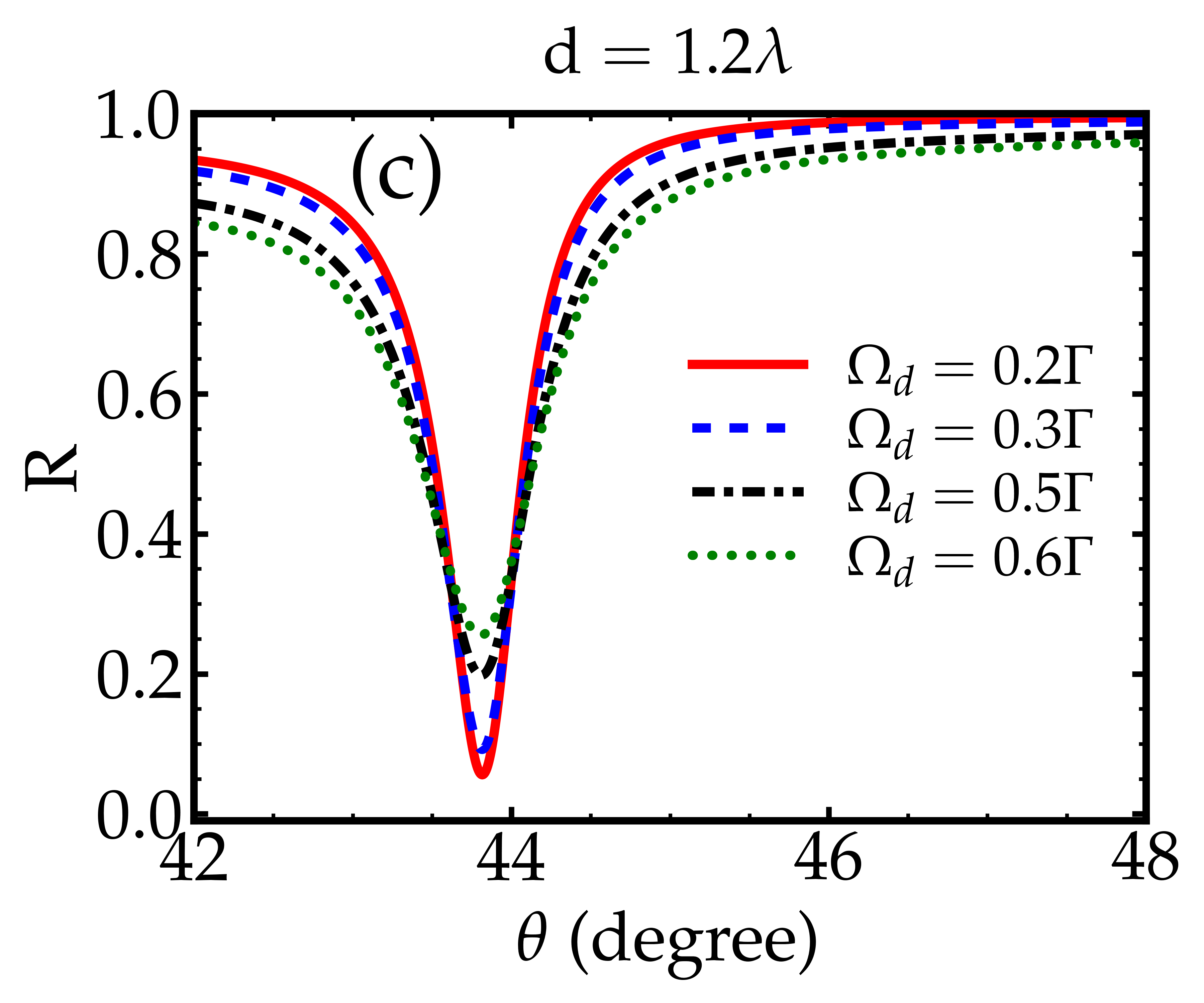} \includegraphics[width = 42mm]{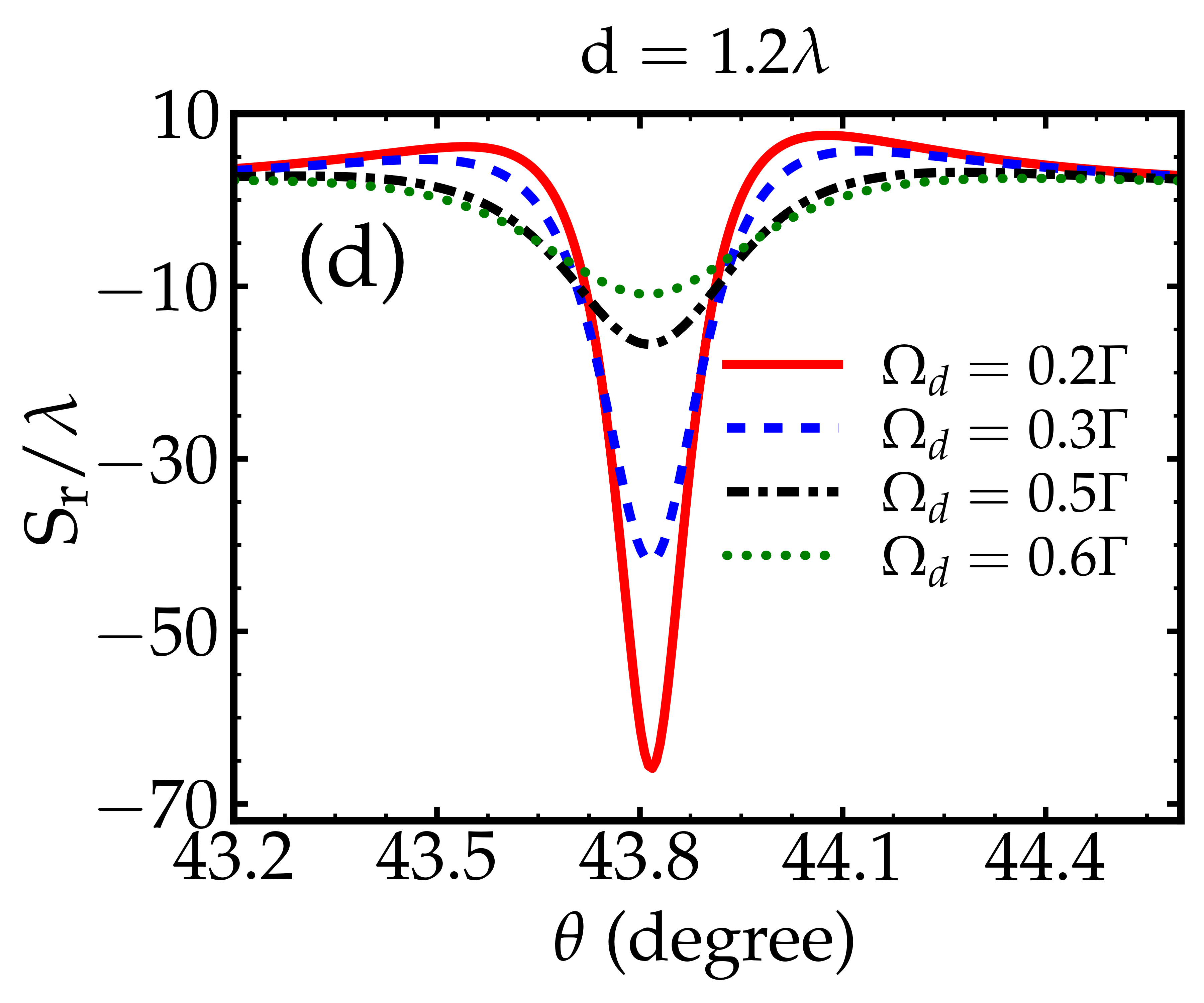}
 \caption[scheme]{Reflectivity and GH shift of the reflected TM-polarized field versus the angle of incidence for different values of Rabi frequency of the driving field $\Omega_d$ while $\Omega_{c} = 2 \Gamma$. In (a) and (b), $d = \lambda$, whereas in (c) and (d), $d = 1.2 \lambda$. Other parameters are $\Gamma_{3} = \Gamma_{4} = \Gamma$, $\beta = 0.04  \Gamma$, $\Delta_{p} = \Delta_{c} = \Delta_{d} = 0$.}
\label{Sub1}
\end{figure}
\section{atomic medium in the gap}
Now, we proceed to investigate the GH shift of the reflected TM-polarized light beam when the air gap between the prism and the metal is doped with the N-type atomic medium. The stationary phase theory expression of the GH shift Eq. \eqref{stationary} shows that GH shift is directly dependent on the phase of the reflection coefficient which can be derived from Eq. \eqref{r}. The reflection coefficient of the layered structure can be manipulated by tuning the permittivity of the atomic medium $\varepsilon_{2} = 1 + \chi$. Therefore, GH shift in this situation can be efficiently controlled by modifying the response of the N-type atomic medium. Because the permittivity $\varepsilon_{2}$ can be actively tuned as discussed in Sec. III, it technically means that the component $k_{z2}$ of the wave vector as well as the Fresnel coefficients $r_{12}$ and $r_{23}$ change for each value of $\varepsilon_{2}$. This obviously leads to different optimal values of the gap thickness and the angles of incidence for which the reflectivity becomes zero. Therefore, the condition $r_{12} + r_{23} \; e^{2i k_{z2} d} = 0$ needs to be solved again for each $\varepsilon_{2}$ in order to determine the corresponding optimal gap thicknesses and the angles that satisfy the zero reflectivity condition.
In order to precisely examine how the optical response of the atomic medium modifies the reflectivity and the GH shift relative to the air-gap case, we use the optimal thickness $d_1$ of the air-gap case as a reference thickness here. Therefore, based on the thicknesses presented in Fig. 2, we consider that the values of the central layer thickness used in this section are $d=\lambda$ and $d=1.2 \lambda$, which are below and above $d_1$, respectively. Then, in the next section, we explore the GH shift around the critical values of $d$ and $\theta$ for each corresponding $\varepsilon_{2}$.
The common parameters we adopt in our calculations are $\Gamma_{3} = \Gamma_{4} = \Gamma$, $\beta = 0.04 \Gamma$, $\varepsilon_{1} = 2.25$, and $\varepsilon_{3} = -13.3+0.883i$.
We consider that the probe, coupling, and driving fields are interacting resonantly with the transitions $\ket{1} \leftrightarrow  \ket{3}$, $\ket{2} \leftrightarrow  \ket{3}$, and $\ket{2} \leftrightarrow  \ket{4}$, respectively. The probe field in this scenario does not encounter any dispersion, i.e., $Re(\chi) = 0$, when propagating in the atomic medium. However, as the absorption of the atomic medium is tunable by modifying the strength of the coupling and driving fields, we show that the GH shift of the reflected TM-polarized field is highly controllable by varying $\Omega_d$ and $\Omega_c$, while the remaining parameters as well as the geometrical characteristics of the layered structure are intact.

In Fig. 3, we plot the reflectivity and GH shift of the reflected TM-polarized field versus the angle of incidence when $d=\lambda$ and $d = 1.2 \lambda$ for different values of Rabi frequency of the driving field $\Omega_d$, while $\Omega_c$ is set to be $2 \Gamma$. As the permittivity of the atomic medium $\varepsilon_{2}$ can be tuned by changing $\Omega_d$, this consequently means that the optimal values of the gap thickness and the angles that satisfy the zero reflectivity condition are different for each $\Omega_d$. Therefore, the two thicknesses used in Fig. 3, which are below and above the optimal air-gap thickness $d_1$ are not necessarily below or above the actual optimal thicknesses of the central layer which correspond to each value of $\Omega_d$. As mentioned before, when the central layer thickness is selected near $d_1$, this allows us to directly see how replacing the air-gap with the atomic medium changes the resonance of the structure and the GH shift behavior. 
Fig. 3(a) shows that the minima of the reflectivities when $\Omega_d  = 0.2\Gamma, 0.3\Gamma,$ and $0.5\Gamma$, are close to zero in comparison to the air-gap case at the same thickness $d= \lambda$. The reason is that solutions of the zero reflectivity condition are modified as the permittivity of the atomic medium $\varepsilon_{2}$ is tuned by $\Omega_d$. Fig. 3(b) shows that the magnitude of the GH shift when $\Omega_d = 0.2\Gamma$ is relatively increased, reaching approximately $120 \lambda$, compared with $\sim 90 \lambda$ when the central layer contains an air, as illustrated in Fig. 2(b) for $d=\lambda$. When Rabi frequency of the driving field is increased to $\Omega_d = 0.3\Gamma$, the GH shift in this case is notably enhanced to $\sim 230 \lambda$. When $\Omega_d$ is increased further, the direction of the GH shift is switched to negative under the effect of the driving field, whereas it is positive for the air-gap case when $d=\lambda$. It is obvious that the response of the atomic medium is sensitive to changing $\Omega_d$, leading to coherent control of the direction and the magnitude of the GH shift.
In order to see how the atomic medium changes the structure resonance response and the behavior of the GH shift when $d>d_1$, we plot the reflectivity and the GH shift for different values of  $\Omega_d$ in Fig. 3(c) and 3(d) when $d = 1.2 \lambda$. The GH shift is negative for all the selected values of $\Omega_d$, preserving the same behavior of the the air-gap when $d>d_1$. As presented in Fig. 3(d), the magnitude of the GH shift is controlled by the Rabi frequency $\Omega_d$. In comparison to the air-gap case where the GH shift is nearly $-110\lambda$ when $d=1.2\lambda$, the atomic medium clearly suppresses the magnitude of the GH shift when $d=1.2\lambda$. It is important to point out that these thicknesses $d=\lambda$ and $1.2\lambda$ do not necessarily represent the optimal thicknesses associated with each $\Omega_d$. \\
When $\Omega_d$ is fixed as shown in Fig. 4, we plot the reflectivity and the GH shift versus the angle of incidence while Rabi frequency $\Omega_c$ is modified for both gap thicknesses $d = \lambda$ and $d =1.2 \lambda$. In this situation, when $d = \lambda$, the GH shift in Fig. 4(b) is negative for all the selected values of $\Omega_c$, with and increasing magnitude as $\Omega_c$ is increased. By comparing this to the air-gap case in the previous section when $d =\lambda$, It is evident that controllable GH shift is accomplished by tuning the Rabi frequency $\Omega_c$. 
When $d =1.2\lambda$, Fig. 4(c) shows that the minima of the plotted reflectivities moved further away from zero, and the GH shift in Fig. 4(d) is obviously suppressed. This indicates that the optimal gap thickness for each $\Omega_c$ in this situation is obviously away from $d =1.2\lambda$. 

\begin{figure}[htbp]
 \centering
 \includegraphics[width=42mm]{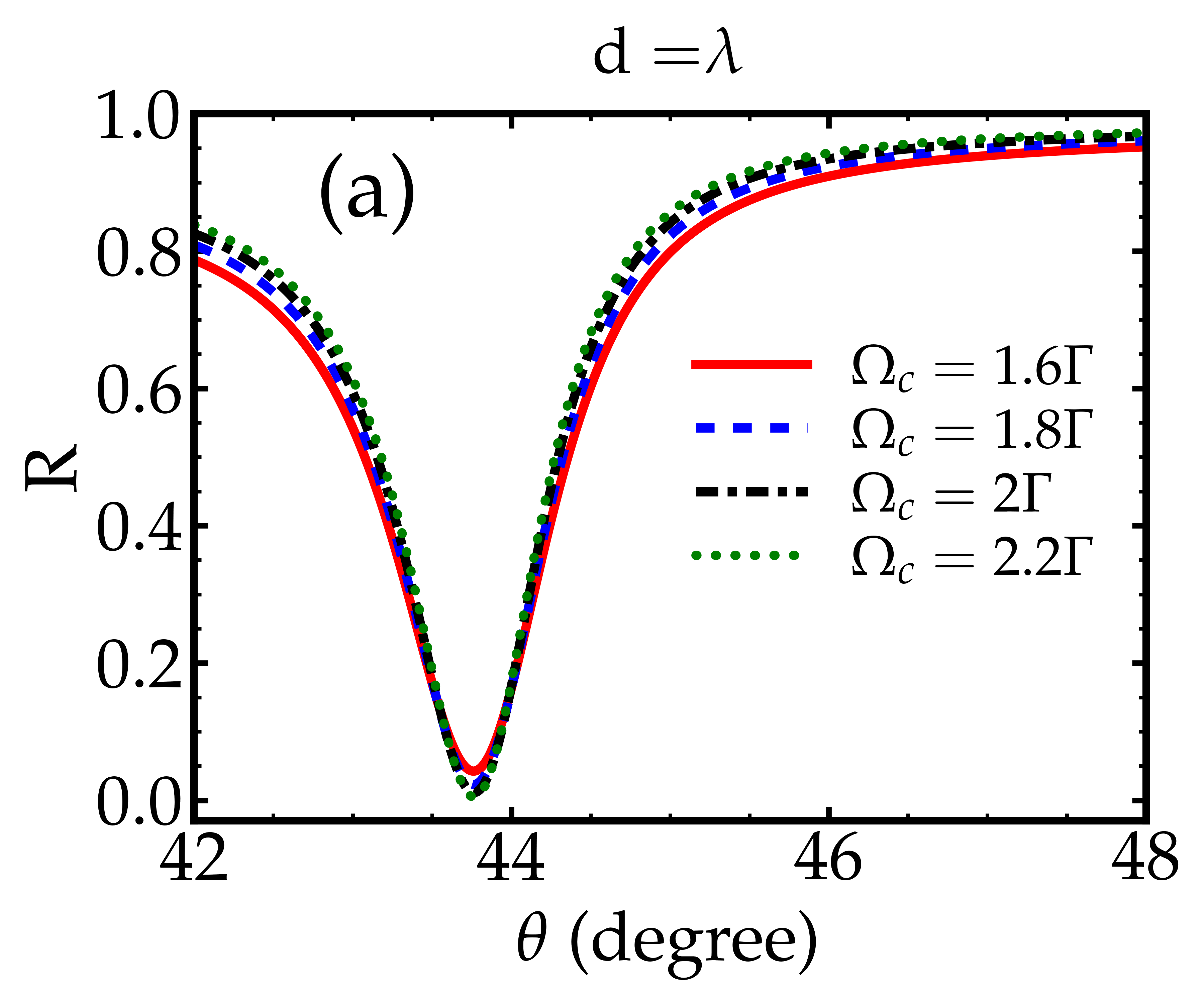} \includegraphics[width = 42mm]{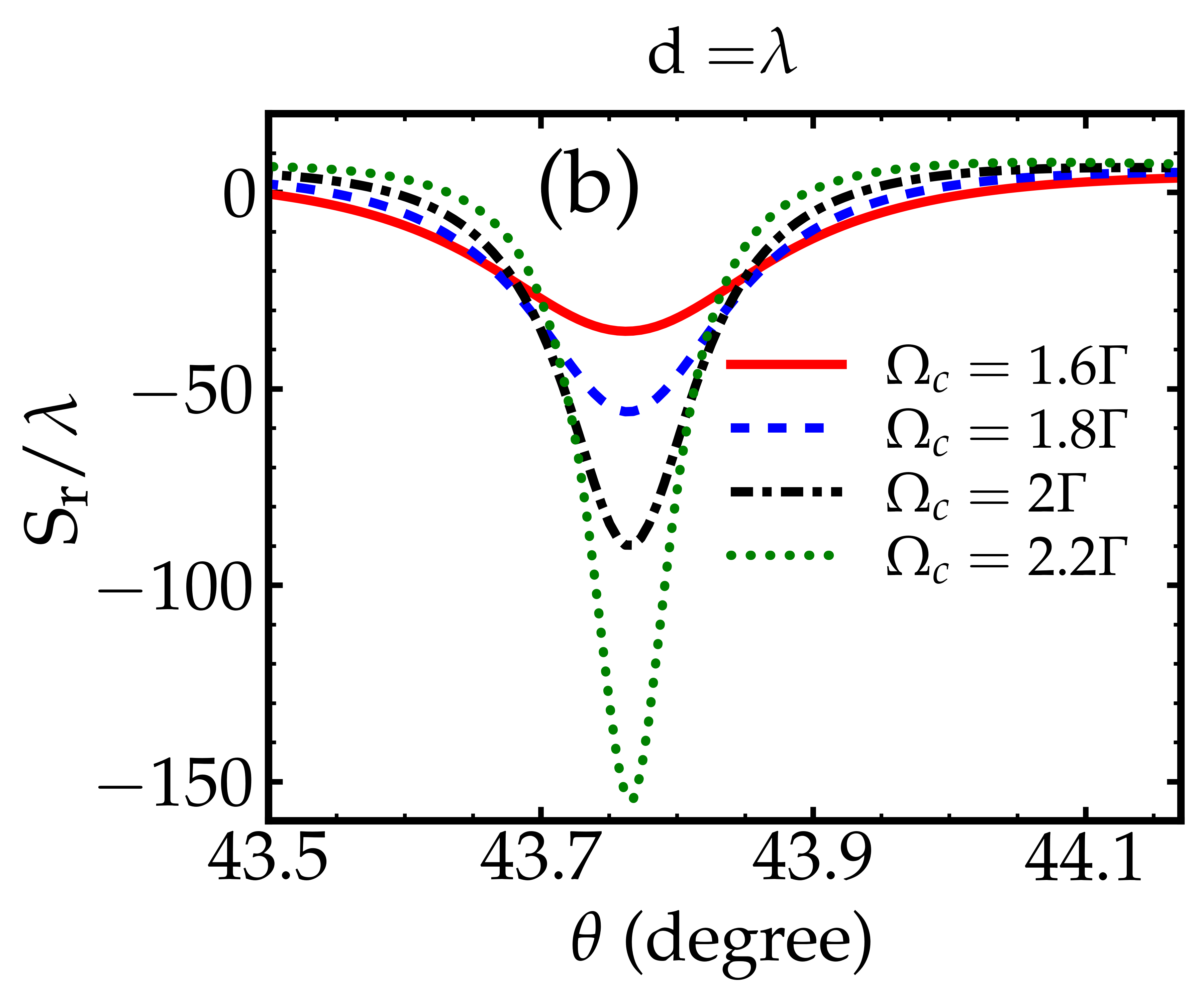} \includegraphics[width=42mm]{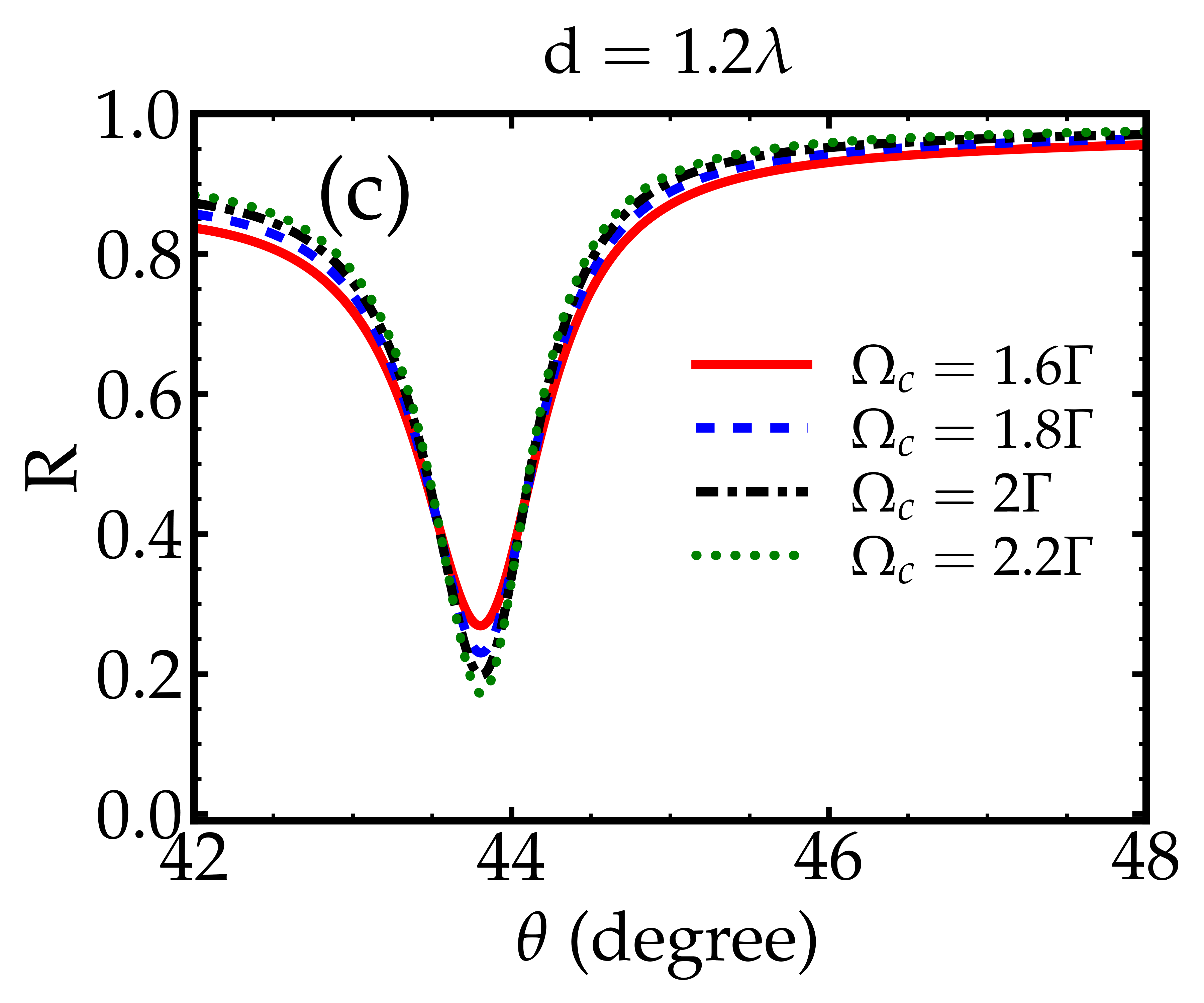} \includegraphics[width = 42mm]{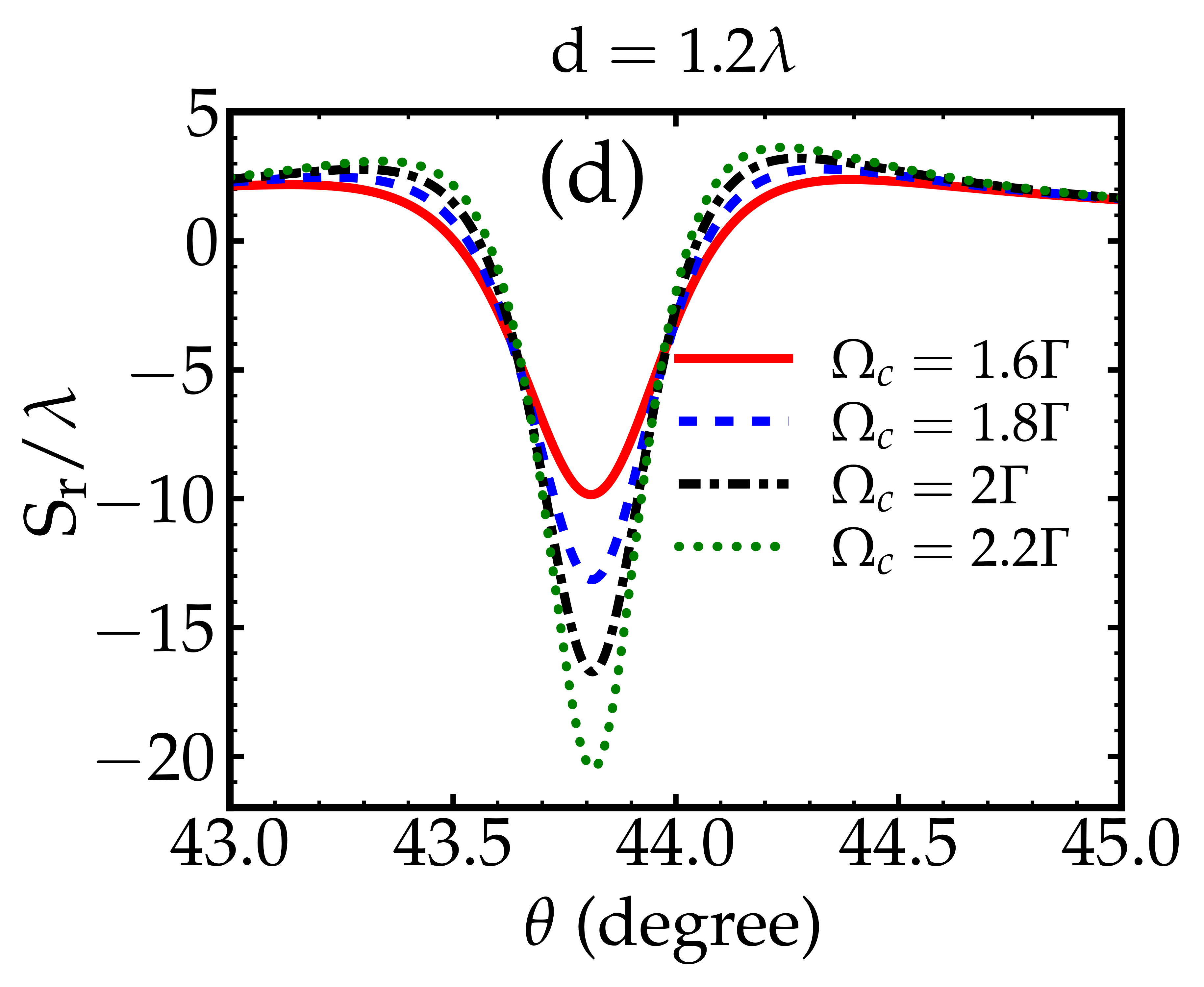}
 \caption[scheme]{Reflectivity and GH shift of the reflected TM-polarized field versus the angle of incidence for different values of Rabi frequency of the coupling field $\Omega_c$ while $\Omega_{d} = 0.5 \Gamma$. In (a) and (b), $d = \lambda$, whereas in (c) and (d), $d = 1.2 \lambda$. Other parameters are the same as in Fig. 3.}
\label{Sub2}
\end{figure}

\section{Dependence of the zero-reflectivity condition on $\varepsilon_{2}$}
The wave vector $k_{z2}$ as well as the Fresnel reflection coefficients $r_{12}$ and $r_{23}$ are explicitly dependent on the permittivity of the medium placed in the central layer $\varepsilon_{2}$. When Rabi frequency of the coupling and driving fields are changed, $\varepsilon_{2}$ is consequently modified, and the optimal values of the central layer thickness as well as the angles at which zero reflectivity occurs are now dependent on $\varepsilon_{2}$. Therefore, the condition $r_{12} + r_{23} \; e^{2i k_{z2} d} = 0$ can be solved numerically for each $\varepsilon_{2}$ in order to find the corresponding $\theta$ and $d$ that satisfy this condition. Thus, we can also see how far the atomic medium shifts the optimal values of the air-gap case, i.e., $\theta_1$ and $d_1$. The Rabi frequencies $\Omega_d$ and $\Omega_c$ are selected based on the results presented in the previous section.
From the illustrated results in Fig. 3, we choose the cases when $\Omega_d = 0.2 \Gamma$ and $\Omega_d = 0.6 \Gamma$. When $\Omega_d = 0.2 \Gamma$, we find that the first zero reflectivity occurs at $\theta \approx 43.8^{\circ} $ and $d \approx 631 \, \text{nm} \simeq 1.07 \lambda$. For the case when $\Omega_d = 0.6 \Gamma$, zero reflectivity occurs at $\theta \approx 43.7^{\circ} $ with $d \approx 530 \, \text{nm} \simeq 0.9 \lambda $. Similarly, we choose from Fig. 4 two cases where $\Omega_d = 0.5 \Gamma$. When $\Omega_c = 1.6 \Gamma$, numerical solution of the zero reflectivity condition gives that the first zero reflectivity is located nearly at $\theta \approx 43.72^{\circ} $ with $d \approx 524 \, \text{nm} \simeq 0.89 \lambda$. When $\Omega_c = 2.2 \Gamma$, the optimal values of the angle and the central layer thickness are $\theta \approx 43.75^{\circ}$ and $d \approx 566 \, \text{nm} \simeq 0.96 \lambda$. It is obvious that the angle of incidence that satisfies zero reflectivity is slightly shifted from the optimal value of the air-gap case $\theta_1 \approx 43.8^{\circ}$. The optimal thicknesses found above indicate substantial deviation from the air-gap optimal thickness $d_1 \approx 1.108 \lambda$ at which the first zero reflectivity occurs.
In Fig. 5, we plot the GH shift versus the angle of incidence in which the thickness of the central layer is chosen to be slightly above and below the optimal thicknesses found above for each $\Omega_d$ and $\Omega_c$. The GH shift near these optimal thicknesses is strongly enhanced and its direction can be switched. When $\Omega_c = 2 \Gamma$ and $\Omega_d = 0.2 \Gamma$, it is clearly noted that the magnitude of the GH shift is enhanced to nearly $10^3$ time the incident wavelength.

\begin{figure}[htbp]
 \centering
\includegraphics[width=42mm]{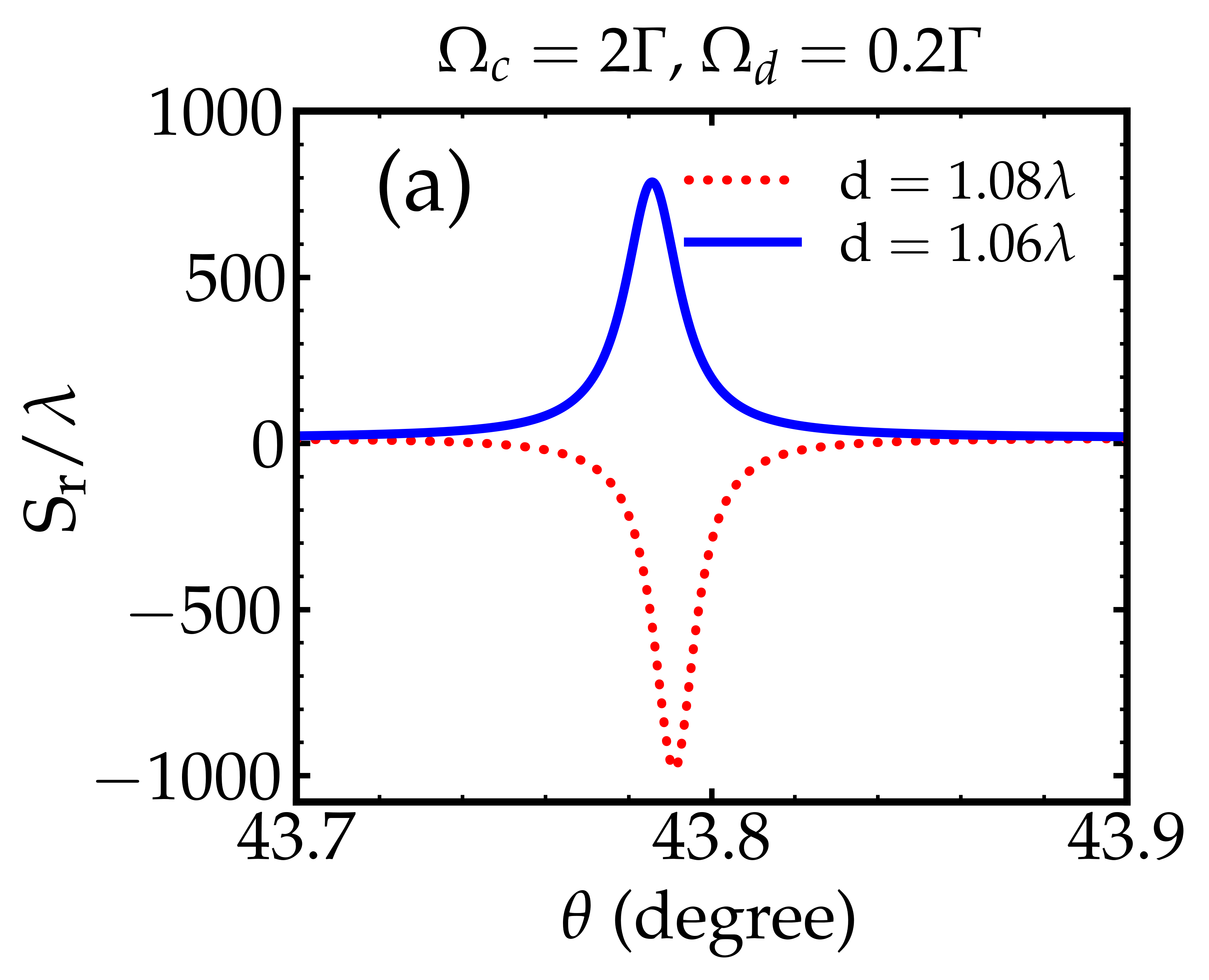} \includegraphics[width = 42mm]{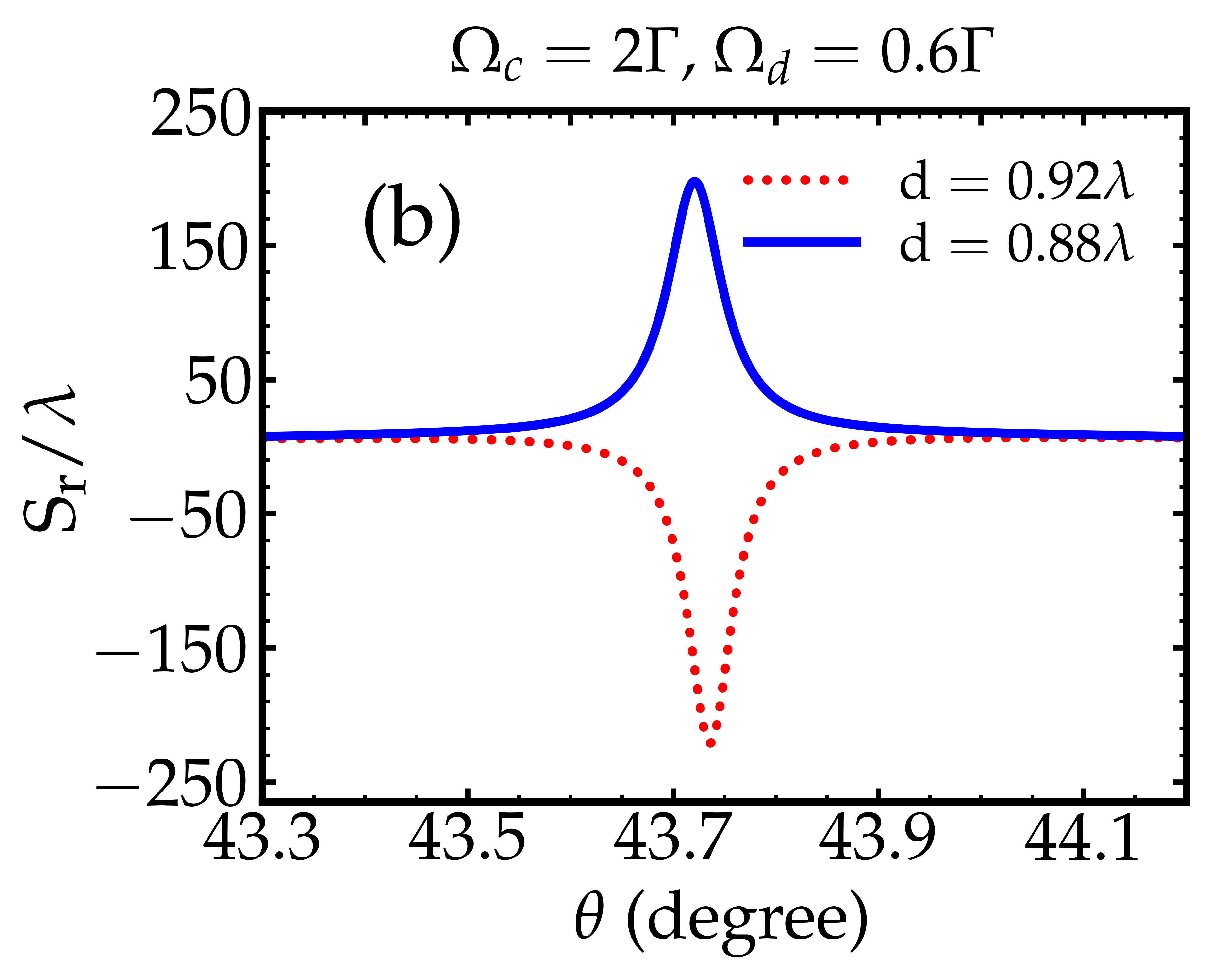}
 \includegraphics[width=42mm]{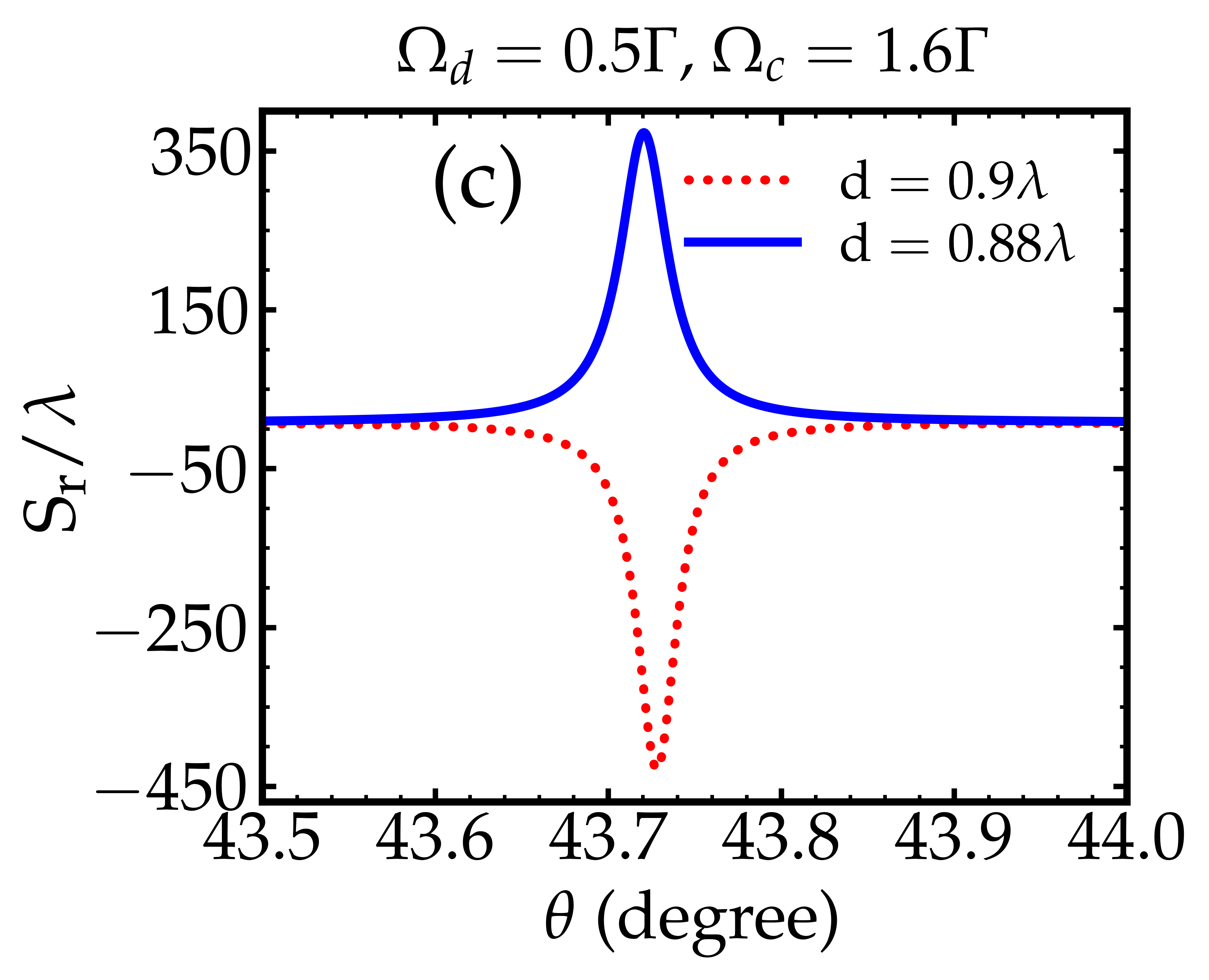} \includegraphics[width = 42mm]{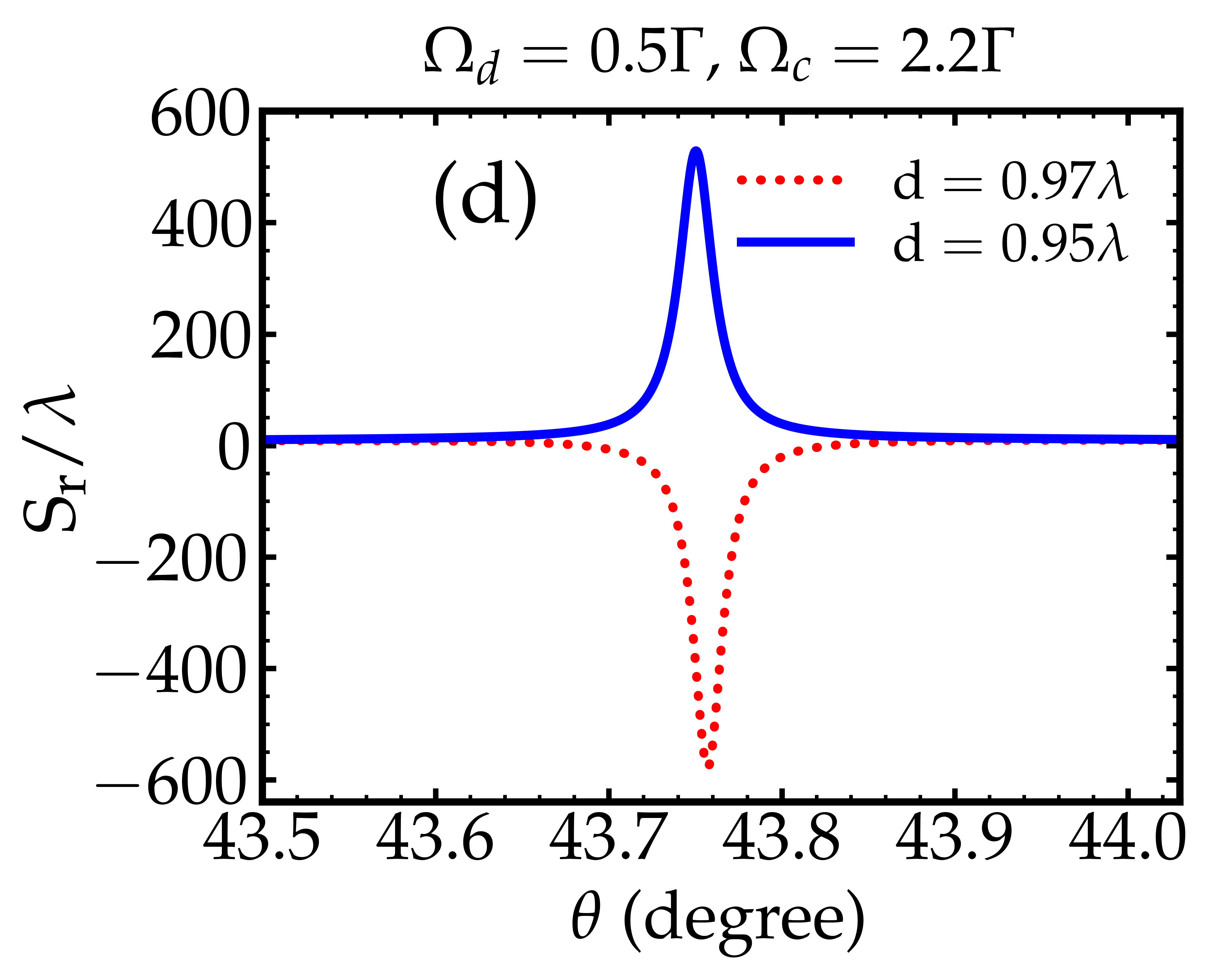}
 \caption[scheme]{GH shift of the reflected TM-polarized field versus the angle of incidence near the optimal thickness of the central layer for different values of Rabi frequency of the coupling and driving fields. In (a), $\Omega_c = 2 \Gamma$ and $\Omega_d = 0.2 \Gamma$. In (b), $\Omega_c = 2 \Gamma$ and $\Omega_d = 0.6 \Gamma$. In (c), $\Omega_d = 0.5 \Gamma$ and $\Omega_c = 1.6 \Gamma$. In (d), $\Omega_d = 0.5 \Gamma$ and $\Omega_c = 2.2 \Gamma$. Other parameters are the same as in Fig .3.}
\label{Sub33}
\end{figure}

For all the cases illustrated in Fig. 5, GH shift is found to be negative when the central layer thickness is above the optimal thickness that corresponds to each $\Omega_c$ and $\Omega_d$, while below these optimal thicknesses, the GH shifts are positive. Therefore, the choice of the gap thickness determines the sign the GH shift.


\section{Conclusion}
We proposed a scheme to manipulate the GH shift of the reflected light beam when a TM-polarized light is incident upon Otto structure, where a coherent atomic medium is placed between the glass prism and the metal film. We showed that the direction and magnitude of the GH shift can be controlled by tuning the atomic medium properties, while the central layer thickness is fixed. As the zero reflectivity condition is dependent on the permittivity of the coherent medium, the required central layer thickness for achieving SPR-related zero reflectivity is consequently different for each $\varepsilon_{2}$. It is found that enhanced GH shifts can be observed near thicknesses that satisfy the zero reflectivity condition for the corresponding permittivity $\varepsilon_{2}$.  

\begin{acknowledgments}
This work is supported by the Ongoing Research Funding program, (ORF-2026-873), King Saud University, Riyadh, Saudi Arabia.
\end{acknowledgments}

\section*{REFERENCES}
\nocite{*}
\bibliography{GH}

\end{document}